\documentstyle[12pt]{article}

\def\Tr{{\rm Tr}\, }
\begin{document}
\hfill {\tt hep-th/0107173}
\vskip0.3truecm
\begin{center}
\vskip 2truecm
{\Large\bf
Horizon Holography
}\\ 
\vskip 1truecm
{\large\bf
Ivo Sachs\footnote{
e-mail: {\tt ivo@theorie.physik.uni-muenchen.de }} and
Sergey N.~Solodukhin\footnote{
e-mail: {\tt soloduk@theorie.physik.uni-muenchen.de}}
}\\
\vskip 0.8truecm
{\it 
Theoretische Physik,
Ludwig-Maximilians Universit\"{a}t, 
\vskip 0.2truecm
Theresienstrasse 37,
D-80333, M\"{u}nchen, Germany}
\vskip 1truemm
\end{center}
\vskip 1truecm
\begin{abstract}
\noindent

A holographic correspondence between horizon data and space-time 
physics is investigated.  We find similarities with the AdS/CFT correspondence, 
based on the observation that the optical metric near the horizon 
describes a Euclidean, asymptotically anti-de Sitter space. 
This picture emerges for a wide class of static space-times with
a non-degenerate horizon, including Schwarzschild black  holes as 
well as de Sitter space-time. We reveal an asymptotic conformal symmetry 
at the horizon. We compute the conformal weights and 2-point functions 
for a scalar perturbation and discuss possible connections with a 
conformal field theory located on the horizon. We then reconstruct 
the scalar field and the metric from the data given on the horizon. 
We show that the solution for the metric in the bulk 
is completely determined in terms of a specified metric on the horizon. From 
the General Relativity point of view our solutions present 
a new class of space-time metrics with non-spherical horizons. 
The horizon entropy associated with these solutions is also discussed.

\end{abstract}
\vskip 1cm
\newpage
\section{Introduction}
\setcounter{equation}0

According to the holographic principle the information about physics in 
space-time would be encoded on a lower dimensional surface, 
the ``holographic screen'' \cite{tHooft, Susskind}.
Given the data on the screen we can reconstruct the events in 
the rest of space-time. A concrete realization of the holographic 
description
was found recently for string theory on anti-de Sitter (AdS) space-time. 
Here 
the holographic screen is the time-like 
boundary of AdS and  the holographic data are given 
in the form of a Conformal Field Theory (CFT) living on the screen 
\cite{Malda,
Gubs, Wit}. According to the AdS/CFT correspondence, these data are then 
enough to reconstruct the physics in the bulk of the 
anti-de Sitter. 
For example, in order to reconstruct the space-time metric  in the bulk 
one has to specify the metric on the screen. This metric in turn couples 
to 
the stress energy tensor of the boundary CFT. 
The quantum expectation value of the dual stress tensor is another piece 
of the CFT data which 
is necessary for the reconstruction.
The explicit 
reconstruction of the metric is given in \cite{CMP}.

In more general situations, when the bulk space-time is not necessarily 
asymptotically anti-de Sitter, but has a horizon the holographic 
information may be encoded on this horizon. 
In this paper we investigate this situation in some detail. 
Concretely, we suggest that horizon holography may be related to 
holography on 
AdS by observing that the optical metric near arbitrary
static (non-degenerate) horizon has the form of direct product of
${{\rm I\!R}}$ (time) and Euclidean anti-de Sitter space. Therefore, some elements 
of
the holographic AdS/CFT dictionary can be transcribed to the present case.
Formulating this dictionary we proceed in two steps. First, we specify
the data on the holographic screen sufficient to reconstruct the physics 
in the bulk. Second, one would like to show that the holographic data 
can be described in terms of some field theory, specifically a conformal 
field theory.
In this paper we mainly concentrate on the first step, that is, 
we analyze in detail the reconstruction of the matter 
fields and metric from the data on the horizon.
Concerning the second step we compute the correlation functions on the 
horizon along the same lines as in AdS/CFT. These correlation functions 
are indeed the same as expected to  arise in some conformal field theory. 
This computation supports the proposed correspondence but further work 
will be 
required to firmly establish such a duality and to reveal the details of 
the underlying CFT. Unfortunately, in the present situation we do not have 
a
string theory formulation of the problem as a guiding principle.

Another motivation for the present work is to understand black hole 
entropy \cite{BH} as arising from some field theory living on the horizon. 
This was carried out explicitly for the $2+1$ dimensional black hole in 
\cite{Carlip0}. In more general situations one then has to investigate 
the general structure of the space-time metric near horizons and to reveal 
the corresponding near-horizon 
asymptotic symmetries. This idea  appeared in recent attempts to 
understand  the black hole entropy in terms of degrees of freedom 
at the horizon \cite{Carlip,SS}. By analyzing a general spherically 
symmetric metric with horizon \cite{SS} or imposing certain fall-off 
condition \cite{Carlip} for the metric coefficients near horizon one 
finds  
the presence of two-dimensional conformal symmetry. This could hint 
to a construction of a Hilbert space for quantum black holes as a 
representation of the conformal symmetry group at the horizon. However, 
in order to incorporate the full dynamics near the horizon it is important 
to analyze the most general class of metrics admitting the interpretation 
in 
terms of horizon and to find the corresponding symmetries. 
In this paper we give up the condition of spherical
symmetry and propose general form for the static metric
with Killing horizon. (A similar form was suggested earlier in
\cite{FS} as a off-shell description of black hole metric near horizon.) 
Remarkably, this metric still can be a solution  to
the Einstein equations. We show this by expanding the metric near 
horizon and observing that all
terms of the expansion are uniquely determined by specifying arbitrary
metric on the horizon (see \cite{Dzhu} for a related discussion for 
spherically symmetric metrics). 
Since the metric on the horizon can be chosen arbitrary the solutions we 
have found 
present a new class of space-times (with non-spherical horizons) of 
General Relativity. Furthermore, the 
asymptotic symmetry 
near the horizon is 
conformal but of different type than observed in \cite{Carlip}, \cite{SS}.

This paper is organized as follows. 
In section $2$ we analyze the optical metric,
exhibit the Euclidean anti-de Sitter space near the horizon and 
compute the conformal weights and $2$-point functions on the horizon. 
In section $3$ we consider the general form of the static metric with
non-degenerate horizon and reveal the conformal asymptotic symmetry.
The reconstruction of the scalar field is considered in section $4$ 
and the reconstruction of the metric
is carried out in section $5$. The entropy associated with the 
non-spherical
horizon of our solutions  is discussed in section $6$. 
The conclusions are then presented in section $7$. 
The reader mainly interested in application of our results to General Relativity
may go directly to sections $3$, $5$ and $6$.

\bigskip

\section{Looking through the optical metric}
\setcounter{equation}0
We begin this section by finding adapted coordinates that exhibit the 
similarities of the geometry near event horizons and Euclidean AdS spaces. 
We then compute some relevant correlation functions in these coordinates 
in the second part. 

\medskip

{\bf 2.1 Uncovering AdS space near the horizon}

\medskip

The key observation important for the present analysis is the universal
appearance of the asymptotically anti-de Sitter space in the optical metric
near the horizon. To see how this comes along let us for simplicity first 
consider  a $(d+2)$-dimensional static, spherically
symmetric metric and transform it to the optical metric
\begin{eqnarray}
&&ds^2=-f(r)dt^2+ f^{-1}(r)dr^2+r^2 d\Omega^2_{(d)} \nonumber \\
&&=f(r)~~ds^2_{opt}~~,
\label{2.1}
\end{eqnarray}
where $d\Omega^2_{(d)}$ is the standard metric on the unit sphere 
$S^d$, in this 
section we will use $\theta$ to denote the various angle coordinates on 
$S^d$.
In terms of the new variable $z=-\int^r f^{-1}(r)dr$ the optical metric 
may be written
in the form
\begin{eqnarray}
&&ds^2_{opt}=-dt^2+ds^2_{spt} \nonumber \\
&&ds^2_{spt}=dz^2+{r^2(z)\over f(z)}d\Omega^2_{(d)}~~.
\label{2.2}
\end{eqnarray}
In the case of interest, when the metric (\ref{2.1}) describes a 
space-time with a non-degenerate
horizon at $r=r_+$ the metric function $f(r)$ has a simple root
at $r=r_+$, $f(r)={2\over \beta_H}(r-r_+)+O(r-r_+)^2$. We interpret 
$2\pi \beta_H$ as the inverse Hawking temperature $T^{-1}_H$ as defined 
with respect to the Killing vector $\partial_t$. 
The horizon surface $\Sigma$ is the $d$-dimensional sphere of radius $r_+$.
For $r$ close to $r_+$ we have
$$
(r-r_+)\sim e^{-{2z\over \beta_H}}~,~~~f(z)\sim  e^{-{2z\over \beta_H}}~~,
$$
so that the spatial part (\ref{3.2}) of the optical metric takes the 
asymptotic form 
\begin{eqnarray}
ds^2_{spt}=dz^2+ C^2 e^{2z\over \beta_H}d\Omega^2_{(d)}~~,
\label{2.3}
\end{eqnarray}
with $C$ is some irrelevant constant. It is not difficult to see that 
the metric (\ref{2.3}) is identical with the asymptotic metric of 
Euclidean anti-de Sitter space. Note that the radius of
the anti-de Sitter space is $\beta_H$, the inverse Hawking temperature of 
the
horizon in the original metric (\ref{2.1}).

In general, the spatial metric (\ref{2.2}) approaches the metric of anti-de 
Sitter only
asymptotically so that in the sub-leading terms the metric will deviate
from that of the anti-de Sitter space. Interestingly, the horizon surface 
$\Sigma$, which is the bifurcation point in the metric (\ref{2.1}), is 
mapped 
to infinity of the anti-de Sitter under the transformation  
$z(r)$. 
This supports the idea that a holographic description in terms of a 
CFT on the boundary of anti-de Sitter
might be applicable to horizons. The dual CFT theory then would be living
on the horizon surface $\Sigma$.

The horizon in metric (\ref{2.1}) can be a black hole horizon or a cosmological
horizon. The cosmological horizon appearing in de Sitter space is of 
particular interest.
In this case the spatial part of the optical metric describes anti-de 
Sitter space  not just
asymptotically but globally.
The metric function in this case  is $f(r)=1-{r^2\over l^2}$ so that 
$\beta_H=l$
($l$ is the radius of de Sitter space). In terms of the new variable
$z=-{l\over 2}\ln {r+l\over r-l}$ we have that
$$
r(z)=l\tanh {z\over l}~,~~{r^2\over f(z)}=l^2\sinh^2{z\over l}
$$
and the spatial part of the optical metric (\ref{2.2})
$$
ds^2_{spt}=dz^2+l^2\sinh^2{z\over l} d\Omega^2_{(d)}~~
$$
is precisely the metric of anti-de Sitter
for all $z$.

\bigskip

{\bf 2.2 Conformal dimensions and correlation functions on the horizon}

\def\ssr{\small r}

\medskip

Consider now scalar perturbations in the background (\ref{2.1})
described by the field equation
\begin{equation}
(\Box -m^2)\phi=0~~,
\label{2.4}
\end{equation}
where $\Box$ is the wave operator for the metric (\ref{2.1}).
This equation can be re-written as a field equation on the background
of the optical metric (2.2). In order to
see that we introduce a new field $\phi_{opt}$ as follows,
$$
\phi=f(r)^{-{d\over 4}} \phi_{opt}~~.
$$
Then the equation for $\phi_{opt}$ reads
\begin{equation}
\Box_{opt}\phi_{opt}=m^2(z)\phi_{opt}~~,
\label{2.5}
\end{equation}
where
\begin{equation}
\Box_{opt}=-\partial_t^2+({f\over r^2})^{d\over 2}\partial_z 
\left( ({r^2\over f})^{d\over 2}\partial_z \right) +{f\over 
r^2}\Delta_{\theta}
\label{2.6}
\end{equation}
is the wave operator in terms of the optical metric (\ref{2.2}), and
\begin{equation}
m^2(z)=m^2f-({d\over 4})^2(f_{,r})^2+{d\over 4}ff_{,rr}+{d^2\over 
4r}ff_{,r}
\label{2.7}
\end{equation}
is an effective, $z$-dependent mass term.

In the near horizon limit, $z\rightarrow \infty$, the spatial part
of the operator (\ref{2.6}) is identical with the Laplace operator 
at asymptotic infinity of Euclidean anti-de Sitter space. 
To continue we split off the time dependence of the optical field, 
$\phi_{opt}= e^{\imath \omega t}
\phi_\omega (z,\theta )$, so that, as $z\rightarrow \infty$,
$\phi_\omega$ satisfies equation
\begin{eqnarray}
&&\Delta_{adS} \phi_\omega=M^2\phi_\omega~~, \nonumber \\
&&M^2=-\omega^2-({d\over 2\beta_H})^2~~.
\label{2.8}
\end{eqnarray}
This is the familiar equation for the scalar perturbations appearing 
in the context of the AdS/CFT correspondence. The unusual feature of
(\ref{2.8}) is the negative sign for the effective mass square term.
This means that the perturbations are tachyonic\footnote{Note, that $M^2$ 
becomes positive when $\omega$ is imaginary. This, in particular, happens 
for quasi-normal modes.}. Asymptotically, 
$$
\phi_\omega \sim \chi_\lambda (\theta ) e^{\lambda z\over 
\beta_H}~,~~z\rightarrow \infty
$$
and from (\ref{2.8}) we find two possible roots for $\lambda_{( \omega )}$
\begin{equation}
\lambda_{( \omega )}^{\pm} =-{d\over 2}\pm \imath \omega \beta_H~~.
\label{2.9}
\end{equation}
In the context of the AdS/CFT correspondence $\lambda_{( \omega )}^\pm$ 
are related to the conformal dimension of a dual operator in the CFT by
\begin{equation}
h_{(\omega )}^\pm=d+\lambda^{( \omega )}_{\pm}={d\over 
2}\pm\imath\omega\beta_H~~.
\label{2.10}
\end{equation}
We see that the conformal weights $h_{(\omega )}^\pm$ are complex. 
So one might worry about unitarity. Note however that, contrary to the 
usual 
AdS/CFT correspondence, the conformal field 
theory in question (if it exists) lives on a Euclidean surface $\Sigma$, 
and does not describe the time evolution of the system.

Near the horizon, the  modes with $\lambda^+_\omega$ and $\lambda^-_\omega$ 
decay in the same 
way so that we can not discard one of them as decaying faster than 
another. Therefore,
the boundary condition near the horizon (or, equivalently, near infinity of 
adS space)
contains both modes,
\begin{equation}
\lim_{z\rightarrow +\infty}\phi_{opt}=
\varphi (t,z,\theta )=e^{-{dz\over 2\beta_H}}\int_{-\infty}^{+\infty}
d\omega \,e^{-\imath \omega t}\left(\varphi^+_\omega (\theta) e^{\imath 
\omega z}+
\varphi^-_\omega (\theta) e^{-\imath \omega z}\right)~~.
\label{2.11}
\end{equation}
In terms of the original scalar field $\phi=e^{{dz \over 2\beta_H}} 
\phi_{opt}$
this is the generic behavior and corresponds to the presence of 
right- and left-moving modes 
which always appear for any field propagating near a horizon. 
In the AdS/CFT picture
$\varphi^\pm_\omega (\theta )$ is dual to an operator 
${\cal O}^\pm_{(-\omega )}$
of conformal dimension $h^\pm_{(\omega )}$. Note that the operators 
${\cal O}^\pm_{(\omega )}$
and ${\cal O}^\mp_{(-\omega )}$ have the same conformal dimension 
$h^\pm_{(\omega )}$.

In order to adjust the prescription of \cite{Wit} for computing the 2-point
functions of dual operators to the present case, let us first consider a 
hypersurface of 
constant $z=Z$ as a boundary, afterwards we will take the 
limit of infinite $Z$.
In this limit the hypersurface approaches the horizon light-cone. 
Considering
(\ref{2.11}) as a boundary condition near the horizon we apply Green's 
formula
\begin{equation}
\phi_{opt}(t,z,\theta )
=\int_{-\infty}^{+\infty} dt'\int_{S^d} d\mu (\theta ')e^{dz'\over \beta_H}
\left(  G\partial_{z'}\varphi (z',t',\theta')-\partial_{z'}G\varphi 
(z',t',\theta')
\right)_{z'=Z}\ ,
\label{2.12}
\end{equation}
where $d\mu (\theta )$ is the integration measure on the sphere $S^d$. 
The boundary function $\varphi (z',t',\theta')$ takes the form 
(\ref{2.11}) and
$$
G(t,t',z,z',\theta , \theta ')=\int_{-\infty}^{+\infty}d\omega 
\,e^{-\imath \omega (t-t')}
G_\omega (z,z',\theta,\theta ')
$$
is Green's function. Near the horizon, $G_\omega$ is a Green function of 
the operator (\ref{2.8})
on anti-de Sitter and we can use the adS expression \cite{SS1}
\begin{equation}
G_\omega \simeq c_\lambda {e^{-(\lambda_\omega +d)z'/\beta_H}\over 
(\cosh {z\over \beta_H}-\sinh {z\over \beta_H} \cos\gamma 
)^{(\lambda_\omega +d)}}~~,
\label{2.13}
\end{equation}
where $\gamma$ is the geodesic distance between two points $\theta$ and 
$\theta '$
on sphere $S^d$. Again, for $\lambda^+_\omega$ and $\lambda^-_\omega$ the 
function 
(\ref{2.13})
decays in the same way for large $z'$. Therefore, the complete Green's 
function is the sum
of two contributions. When both points $z$ and $z'$ are close to the 
boundary of adS
we have
\begin{eqnarray}
&&G_\omega=e^{-{d\over 2 \beta_H}(z+z')}\left(G^-_\omega (\theta, 
\theta')e^{\imath\omega (z+z')}+
G^+_\omega (\theta, \theta')e^{-\imath\omega (z+z')}\right)~~, \nonumber \\
&&G^-_\omega (\theta, \theta')={c^+_\omega \over 
(1-\cos\gamma)^{h^+_\omega}}~,~~
G^+_\omega (\theta, \theta')= {c^-_\omega \over 
(1-\cos\gamma)^{h^-_\omega}}~~.
\label{2.14}
\end{eqnarray}
In this limit Green's formula (\ref{2.12}) indeed gives the function which 
satisfies
the boundary condition (\ref{2.11}). It also leads to non-local relations 
between
$\varphi^+_\omega$   and  $\varphi^-_\omega$
\begin{eqnarray}
&&\varphi^+_\omega(\theta )=-2\imath \omega \int_{S^d}d\mu 
(\theta')G^-_\omega (\theta , \theta ')
\varphi^-_\omega(\theta' ) \nonumber \\
&&\varphi^-_\omega(\theta )=2\imath \omega \int_{S^d}d\mu 
(\theta')G^+_\omega (\theta , \theta ')
\varphi^+_\omega(\theta' )~~.
\label{*1}
\end{eqnarray}
We note in passing that, as a consequence of (\ref{2.12}) and (\ref{2.15}) 
the components of the Green's function 
(\ref{2.14}) should satisfy the consistency condition\footnote{
This relation is valid for kernels (\ref{2.14}) with any $h^+$ and $h^-$ 
such that
$h^++h^-=d$.  Most easily it can be proved in flat space (sphere of 
infinite radius)
by doing the Fourier transformation of the kernels. 
Furthermore, in the 
particular 
case where $h^+={d\over 2}+{1\over 2}$ and $h^-={d\over 2}-{1\over 2}$, 
the kernels $G^-$ and $G^+$ are respectively Dirichlet and Neumann
correlation function \cite{SS1}. 
The proof of (\ref{*2}) in this case was explained to us by A. Barvinsky. 
}
\begin{equation}
\int_{S^d}d\mu (\theta'')G^-_\omega (\theta , \theta '')G^+_\omega (\theta 
'' , \theta ')=
{1\over 4\omega^2} \delta^{(d)}(\theta ,\theta ' )~~.
\label{*2}
\end{equation}
Now, in order to compute the 2-point functions of dual operators we have 
to evaluate 
the boundary action  $W_B$ on the solution (\ref{2.12}). In terms of 
$\phi_{opt}$ and the optical metric one has 
\begin{eqnarray}
W_B[\varphi ]&=&{1\over 2}\int_{-\infty}^{+\infty} dt \int_{S^d}d\mu 
(\theta )
\left(\phi_{opt}e^{{dz\over 2\beta_H}}\partial_z (e^{{dz\over 2\beta_H}}
\phi_{opt})\right)_{z=Z}\\
&=&-{1\over 2}\int_{S^d}d\mu(\theta )\, \int_{-\infty}^{+\infty} d\omega \ 
i\omega\left(\varphi^+_{\omega }(\theta )\varphi^+_{-\omega }(\theta )-
\varphi^-_{\omega }(\theta )\varphi^-_{-\omega }(\theta )\right) + 
\nonumber~~\cdots,
\label{2.15}
\end{eqnarray}
where the periods stand for contact terms which we will discard in the 
following. 
The exponent of $W_B[\varphi ]$ should be compared with the 
generating functional 
$$
\left\langle\exp\left[\int_{S^d} d\mu (\theta )\int_{-\infty}^{+\infty}d\omega
\left(\varphi^+_{-\omega 
}(\theta )
{\cal O}_\omega^+(\theta )+\varphi^-_{\omega}(\theta){\cal O}_{-\omega}^-
(\theta )\right)\right]\right\rangle
$$
on the CFT side. Using (\ref{2.14}) we then read off the correlation 
functions of the dual 
operators 
\begin{eqnarray}
&&<{\cal O}_\omega^+{\cal O}_{-\omega}^- >=\omega^2 G^-_\omega\propto {1\over 
(\sin {\gamma\over 2})^{2h^+_\omega}} \nonumber \\
&&<{\cal O}_\omega^-{\cal O}_{-\omega}^+ >=\omega^2 G^+_\omega\propto {1\over 
(\sin {\gamma\over 2})^{2h^-_\omega}} ~~,
\label{2.16}
\end{eqnarray}
which are precisely the correlation functions expected arise on 
sphere $S^d$ for operators with dimension $h^+_\omega$ and $h^-_\omega$ 
respectively.

\medskip

{\bf Remarks:}

\medskip

{\it 1. Extra boundary}.
In the above computation of correlation functions
we neglected the presence of boundaries different from the horizon.
An extra boundary at infinity appears for instance when space-time 
is 
asymptotically flat or anti-de Sitter. Then the computation should 
also produce 2-point functions between different boundary components.

{\it 2. Hawking radiation}. 
We should also notice the appearance of
the inverse Hawking temperature $\beta_H$ in combination with $\omega$ in 
our formulas.
Since $\omega$ is the energy of the perturbation it is tempting 
to speculate that the dual operators ${\cal O}_{(\omega )}^\pm$ 
with conformal dimension $h_{(\omega )}^\pm$ could enter in a dual 
description of the Hawking radiation in terms of a CFT. However, we 
do not explore this possibility any further in this paper leaving this 
issue to a future work.

{\it 3. Similarity to 't Hooft's S-matrix}. When the horizon is 
two-dimensional ($d=2$ in the above formulas)
we may introduce the function $f_0(\theta , \theta ')=
-\ln\sin^2 {\gamma\over 2}$
which, for $\theta \neq \theta '$, is a solution to 
the equation $\Delta_{S^2}f_0={1\over 2}R$
($R$ is curvature scalar of $S^2$). For $\theta$ 
close to $\theta'$ it behaves as usual Green's
function in two dimensions, $f_0(\theta ,\theta')\simeq 
-\ln |\theta -\theta '|^2$. Then the kernels
$G^\pm$ can be written as
$$
G^\pm=\exp \left( (1\pm \imath \omega \beta_H)f_0(\theta ,\theta') \right) 
~~.
$$ 
For Schwarzschild black hole we have 
$\omega \beta_H=4GM\omega$, where $M$ is the mass of the 
black hole and $\omega$
is the energy of a particle falling into black hole.
The imaginary part of the exponent in $G^\pm$ 
resembles the phase appearing in the two-particle amplitude for the 
gravitational scattering in the eikonal approximation \cite{GtH}, \cite{VV}
$$
S_{12}=\exp \left(2\imath G p_1^-p_2^+ f(y_1,y_2)\right)~~, 
$$
$$
f(y_1,y_2)=-\ln |y_1-y_2|^2~~,
$$
where $p^\pm_i$ and $y_i$ are the momentum and transversal 
coordinate of particle $i$.
The horizon sphere with angle coordinates $\theta$ indeed plays 
a role of a transversal
space for a particle falling into a black hole. Moreover, according 
to (\ref{*1})
$G^\pm$ relate the in-going and out-going modes at the horizon. 
Therefore, this 
similarity may not be accidental and deserves a deeper analysis.

{\it 4. de Sitter space}.
Our computation of the correlation functions is, in particular, 
applicable to de Sitter space. The 2-point
functions (\ref{2.16}) are then defined on the cosmological horizon. 
Recently, there was some interest in quantum gravity and string theory on 
de Sitter space \cite{Wit1}, \cite{Str}. In particular, some form of  
dS/CFT duality was proposed in \cite{Str}.  
The conformal field theory in this proposal lives on a Euclidean space 
which is a hypersurface of infinite past in de Sitter space. 
The computation of the 2-point functions in \cite{Str} for the CFT defined 
on 
infinite past of de Sitter has some similarities with our computation in 
this 
section. In particular, the necessity of more general boundary conditions 
including both modes with $h^+$ and $h^-$ 
also arises there. In spite of these similarities the two CFT's are defined
in different space-time regions. It would be interesting to understand 
the relation between these two approaches.

\bigskip

\section{General metric and asymptotic symmetries}
\setcounter{equation}0

What we have found in the previous section is that the optical metric of a 
spherically symmetric space-time with horizon 
is a direct product of time and a Euclidean space which near the horizon 
asymptotically  approaches anti-de Sitter
space. As we will show, this observation applies in fact 
to a more general class of metrics  with a horizon. 
Motivated by recent study
\cite{FG}, \cite{GL}, \cite{HS}, \cite{CMP}
on asymptotically anti-de Sitter spaces we consider a general static 
ansatz of form
\begin{eqnarray}
&&ds^2=e^{\sigma (x,\rho )} 
\left(-dt^2 +{l^2d\rho^2\over 4\rho^2}+{1\over \rho}g_{ij}(x,\rho 
)dx^idx^j \right)~~, \nonumber 
\\
&&\sigma (x, \rho )=\ln \rho +\sigma_{(0)} (x) + O(\rho )~~, 
\label{3.1}
\end{eqnarray}
that is the function
$e^\sigma$ has a simple root at $\rho=0$ corresponding to the location of 
the horizon.
For the spatial part of the 
optical metric
in (\ref{3.1}) the horizon surface defined by $\rho=0$ is the boundary at 
infinity. Provided the function $g_{ij}(x,\rho )$ is analytic near 
$\rho=0$ and
approaches $g_{(0)\,ij}(x)$ when $\rho=0$, we find that the spatial 
metric 
in (\ref{3.1})
indeed describes asymptotically anti-de Sitter space as $\rho$ goes to 
zero. 
The parameter $l$ in (\ref{3.1}) 
determines the radius of this anti-de Sitter. The inverse Hawking 
temperature 
for the horizon as defined with respect to $\xi_t=\partial_t$ in the 
metric (\ref{3.1}) reads
\begin{equation}
\beta^{-1}_H=l^{-1}\rho \partial_\rho \sigma |_{\rho=0}~~.
\label{3.2}
\end{equation}
Hence, if $e^\sigma\sim \rho$ for small $\rho$, then $\beta_H=l$. 
The metric (\ref{3.1}) is the general form for a static 
metric 
with a Killing horizon. This form was proposed earlier in \cite{FS} as
 off-shell form of the metric near an arbitrary, static black hole horizon.
In the next section we will prove that (\ref{3.1}) is, in fact, valid also 
on-shell by determining the 
functions $\sigma (x,\rho )$ and $g_{ij} (x, \rho )$ which 
solve the $(d+2)$-dimensional 
Einstein equations.

First, let us however analyze the ($d+2)$-dimensional 
diffeomorphisms which preserve  the form (\ref{3.1}).
This is again motivated by the analysis done for anti-de Sitter space 
where the conformal transformations in the boundary originate from 
a subset of diffeomorphisms in the bulk \cite{Brown,ISTY} (see also \cite{CMP}). 
In general the diffeomorphisms $\xi$ can be functions of 
$x,~\rho$ and time $t$.
However, at present we restrict ourselves to the static case and 
assume 
that $\partial_t\xi^\rho=\partial_t\xi^i=0$. This, however, does not mean 
that $\xi^t$ can not depend on $t$. The most general 
diffeomorphism leaving invariant the form (\ref{3.1}) is then of the form 
\begin{eqnarray}
&&\xi^t=qt+s ~~,~~~\xi^\rho= \rho~\alpha (\rho, x)~~,~~~\xi^i=a^i(x,\rho 
)~~, \nonumber \\
&&\alpha (\rho , x)= q \ln\rho +\alpha_0(x)
\label{3.4}
\end{eqnarray}
where $q,s$ are constant and $a^i(x,\rho )$ is subject to the constraint 
\begin{equation}
\partial_\rho a^i (x, \rho )=-{1\over 4} g^{ij}(x,\rho ) \partial_j 
\alpha_0 (x)\ ,
\label{3.6}
\end{equation}
similar to the equation arising in the analysis of diffeomorphisms in 
anti-de Sitter \cite{ISTY}. Under these diffeomorphisms the functions 
$\sigma (x,\rho )$
and $g_{ij}(x,\rho )$ transform as follows
\begin{equation}
{\cal L}_\xi \sigma=a^i(x,\rho )\partial_i \sigma +
\alpha \rho\partial_\rho \sigma +2\rho\partial_\rho \alpha~~,
\label{3.7}
\end{equation}
\begin{equation}
{\cal L}_\xi g_{ij}=\nabla_i(g_{jk}\xi^k)+\nabla_i(g_{ik}\xi^k)-\alpha 
(g_{ij}-\rho\partial_\rho g_{ij})
-2\rho\partial_\rho \alpha g_{ij}~~.
\label{3.8}
\end{equation}
The covariant derivative in (\ref{3.8}) is with respect to the metric 
$g_{ij}(x,\rho )$. The metric induced  on a surface of constant 
$\rho$ ($\rho=0$ is the horizon) is 
$\gamma_{ij}(x,\rho) =e^\sigma {1\over \rho} g_{ij}$. 
Under the diffeomorphisms (\ref{3.4}) it then transforms as
\begin{equation}
{\cal L}_\xi \gamma_{ij}=\nabla^{(\gamma 
)}_i(\gamma_{jk}a^k)+\nabla^{(\gamma )}_i(\gamma_{ik}a^k)
+\alpha \rho\partial_\rho \gamma_{ij}~~,
\label{3.9}
\end{equation}
where $\nabla^{(\gamma )}_i$ is the covariant derivative with respect to 
$\gamma_{ij}(x,\rho )$.

For $q=0$ the diffeomorphisms (\ref{3.4}) are similar to the those found 
in 
\cite{ISTY} for the case of AdS. Since $\alpha (x)$ is not function of 
$\rho$ the last term in (\ref{3.7}) and (\ref{3.8}) disappears. 
On the horizon surface ($\rho=0 $) the transformations of this  type 
act as conformal transformations, 
\begin{eqnarray}
&&{\cal L}_{\xi}\sigma_{(0)}(x)=\alpha_0(x) \nonumber \\
&&{\cal L}_{\xi}g_{(0)\,ij}(x)=-\alpha_0(x)g_{(0)\,ij}(x)~~. 
\label{Weyl}
\end{eqnarray}
On the other hand, for the induced metric $\gamma_{ij}(x, \rho )$
we find from (\ref{3.9}) that
\begin{equation}
{\cal L}_{\xi}\gamma_{ij}(x)=0~~,
\label{3.14}
\end{equation}
i.e. the induced metric on the horizon does not change. 
Thus, in spite of the fact that the parameter $\alpha_0(x)$ is function on 
the $d$-dimensional
horizon surface $\Sigma$, the transformation (\ref{3.4}) (with $q=0$) 
effectively 
acts as Weyl transform
in the two-dimensional  $(t,\rho )$ subspace orthogonal to $\Sigma$. 
The induced metric
on $\Sigma$ remains invariant under this.

The remaining diffeomorphisms in (\ref{3.4}), parametrized by $q$ 
lead to the transformations 
\begin{eqnarray}
&&{\cal L}_{\xi} \sigma=q(2+\rho\ln\rho \partial_\rho \sigma ) \nonumber \\
&&{\cal L}_{\xi} g_{ij}=q(-2 g_{ij}-\ln\rho g_{ij}+\rho\ln\rho 
\partial_\rho 
g_{ij}) 
\nonumber \\
&& {\cal L}_{\xi} \gamma_{ij}=q \ln\rho\partial_\rho \gamma_{ij}
\label{3.10}
\end{eqnarray}
These infinitesimal transformations can be integrated to the finite 
transformations
\begin{eqnarray}
&&\sigma (x,\rho )\rightarrow \sigma (x,\rho^{e^q}) +2q \nonumber \\
&&g_{ij}(x, \rho )\rightarrow e^{-2q} g_{ij}(x, \rho^{e^q} )\nonumber \\
&&\gamma_{ij}(x, \rho )\rightarrow \gamma_{ij}(x,\rho^{e^q} )
\label{3.11}
\end{eqnarray}
and results in simply replacing $\rho$ by $\rho^{e^q}$ in these functions. 
The interpretation of this transformation is that it changes the 
definition 
of the Hawking temperature. Indeed we find from (\ref{3.2}) and 
(\ref{3.11}) 
that
\begin{equation}
\beta^{-1}_H\rightarrow e^q ~ \beta^{-1}_H~~.
\label{3.12}
\end{equation}

\bigskip

\section{Reconstruction of the scalar field}
\setcounter{equation}0

In this section we consider a scalar perturbation on a general background 
with horizon. The scalar field equation
\begin{equation}
(\Box-m^2)\Phi=0
\label{4.1}
\end{equation}
is the same as in Section 2 but now the background is given 
by (\ref{3.1}). To continue we suppose that 
the functions $\sigma (x,\rho )$ and $g_{ij}(x, \rho )$ in the metric 
(\ref{3.1}) are given in terms of an expansion in $\rho$, i.e. 
\begin{eqnarray}
&&\sigma(x,\rho )=\ln \rho +\sigma_{(0)}(x)+\sigma_{(1)}(x)\rho +... 
\nonumber \\
&&g_{ij}(x,\rho )=g_{(0)\,ij}(x)+g_{(1)\,ij}(x)\rho +...
~~.
\label{4.2}
\end{eqnarray}
Our purpose in this section is to understand what data one should specify 
on horizon in order to 
reconstruct the field everywhere in the bulk.

At fixed energy $\omega$  the perturbation splits on two sectors
\begin{equation}
\Phi_\omega = 
e^{\imath \omega t}\left(\rho^{\imath \omega\beta_H\over 2}\phi_\omega 
(x,\rho )+
\rho^{-{\imath \omega\beta_H\over 2}}\psi_\omega (x,\rho )\right)~~,
\label{4.3}
\end{equation}
which are the $h^+_\omega$- and $h^-_\omega$-sectors discussed in Section 
2.
They describe right- and left-moving waves at the horizon. For the 
functions $\phi_\omega (x,\rho )$
and $\psi_\omega (x,\rho )$ we then have the following expansion
\begin{eqnarray}
&&\phi_\omega (x,\rho )=\sum_{k=0}^\infty \phi^{(k)}_\omega (x)\rho^k 
\nonumber \\
&&\psi_\omega (x,\rho )=\sum_{k=0}^\infty \psi^{(k)}_\omega (x)\rho^k~~.
\label{4.4}
\end{eqnarray} 
The two sectors decouple in the field equation (\ref{4.1}).
Therefore, we will restrict the consideration to the $\phi_\omega$-sector. 
The
analysis for $\psi_\omega$ is similar.
At fixed $\omega$ the equation on $\phi_\omega (x,\rho )$ reads
\begin{eqnarray}
&&\imath \omega \beta_H \phi_\omega \left( 
d(\rho\partial_\rho\sigma-1)+\rho \Tr (g^{-1}\partial_\rho g) \right) 
\nonumber \\
&&+\rho\partial_\rho \phi_\omega \left(4+\imath 4\omega 
+2d(\rho\partial_\rho\sigma-1)
+2\rho \Tr (g^{-1}\partial_\rho g)\right) \nonumber \\
&&+4\rho^2\partial_\rho^2 \phi_\omega +\rho \nabla^2_g \phi_\omega +{d\over 
2}\rho g^{ij}\partial_i\sigma
\partial_j\phi_\omega -m^2e^\sigma \phi_\omega=0~~.
\label{4.5}
\end{eqnarray}
Substituting the expansion (\ref{4.4}) into equation (\ref{4.5}) and 
assuming
that all coefficients in the expansion  (\ref{4.2}) are known we solve eq. 
(\ref{4.5}) perturbatively at each order in
$\rho$. It proves that all terms of the expansion (\ref{4.4}) are uniquely 
determined provided
the first term $\phi^{(0)}_\omega (x)$ is given. The same is true also for 
the
$\psi_\omega$-sector where one has to fix the function $\psi^{(0)}_\omega 
(x)$ 
on the horizon.
The first two terms in (\ref{4.4}) are determined as follows
\begin{eqnarray}
\phi^{(1)}_\omega (x)&=&-{1\over 4(1+\imath \omega\beta_H )} [\imath 
\omega\beta_H (d\sigma_{(1)}+
\Tr g_{(1)})\phi^{(0)}_\omega  \nonumber \\
&&+\nabla^2_{(0)}\phi^{(0)}_\omega +{d\over 
2}g^{ij}_{(0)}\partial_i\sigma_{(0)}\partial_j
\phi^{(0)}_\omega -m^2e^{\sigma_{(0)}}\phi^{(0)}_\omega ]\ ,
\label{4.6}
\end{eqnarray}
\begin{eqnarray}
\phi^{(2)}_\omega (x)&=&-{1\over 8}(d\sigma_{(1)}+\Tr 
g_{(1)})\phi^{(1)}_\omega 
-{\imath \omega\beta_H \over 8(2+\imath \omega 
\beta_H)}(2d\sigma_{(2)}+2\Tr g_{(2)}-\Tr g^2_{(1)}) \nonumber \\
&&-{1 \over 8(2+\imath \omega \beta_H)}[\nabla^2_{(0)}\phi^{(1)}_\omega 
+\nabla_i (g^{ij}_{(1)}
\partial_j\phi^{(0)}_\omega )+{1\over 2}\partial_i\Tr g_{(1)} 
\partial^i\phi^{(0)}_\omega  \\
&&+{d\over 2}(\partial_i\sigma_{(0)}\partial^i \phi^{(1)}_\omega 
+\partial_i\sigma_{(1)}\partial^i 
\phi^{(0)}_\omega-g^{ij}_{(1)}\partial_i\sigma_{(0)}\partial_j\phi^{(0)}_\omega 
) 
-m^2e^{\sigma_{(0)}}(\phi^{(1)}_\omega +\sigma_{(1)}\phi^{(0)}_\omega )]\ .\nonumber
\label{4.7}
\end{eqnarray}
At higher orders the expressions become more complicated but the general 
structure of 
equation for $k$-th coefficient is always of the form\footnote{One can see 
that 
for complex $\omega$ the has a pole at $\Im\omega =-  2\pi T_H k$, where $k$ is
integer. It is tempting to relate it with the quantization of imaginary 
part of
frequency $\omega$ for quasi-normal modes. This, however, needs a more 
careful
analysis.} 
\begin{equation}
\phi^{(k)}_\omega k(\imath \omega \beta_H +k) 
+X_{(k)}[\phi^{(k-1)}_\omega, \phi^{(k-2)}_\omega, ...]=0\ ,
\label{4.8}
\end{equation}
where $X_{(k)}$ is a polynomial in $\phi^{(p)}_\omega,\,\,p<k$ and their 
derivatives. Thus, the coefficient $\phi^{(k)}_\omega (x)$ is completely 
determined by the previous coefficients
$\phi^{(k-1)}_\omega$, $\phi^{(k-2)}_\omega$, ..., $\phi^{(0)}_\omega$
and ultimately by the function $\phi^{(0)}_\omega (x)$. 
A separate analysis is needed for zero-energy ($\omega=0$) modes. In this 
case some terms
in eq.(\ref{4.5}) disappear. Moreover, instead of the expansion (\ref{4.3}), 
(\ref{4.4}) we have
\begin{equation}
\Phi_0=(\phi_{(0)}(x)+\rho\phi_{(1)}(x) +...)+\ln\rho~ 
(\psi_{(0)}(x)+\rho\psi_{(1)}(x) +...) ~~. 
\label{4.9}
\end{equation}
This time, $\phi$- and $\psi$- sectors couple to each other in the field 
equation. 
Nevertheless, specifying 
two functions $\phi_{(0)}(x)$ and $\psi_{(0)}(x)$ on the horizon 
completely determines all terms in the expansion (\ref{4.9}). In the first 
order in $\rho$ we find
\begin{eqnarray}
\psi_{(1)}(x)&=&-{1\over 4}\left( \nabla^2_{(0)}\psi_{(0)}+{d\over 
2}g^{ij}_{(0)}\partial_i\sigma_{(0)}
\partial_j\psi_{(0)}-m^2e^{\sigma_{(0)}}\psi_{(0)} \right)~~, \nonumber \\
\phi_{(1)}(x)&=&-{1\over 3}\psi_{(1)}(x)-{1\over 6} (d\sigma_{(1)}+\Tr 
g_{(1)})\psi_{(0)}(x) 
\nonumber \\
&&+{1\over 12 }\left(\nabla^2_{(0)}\phi_{(0)}+{d\over 
2}g^{ij}_{(0)}\partial_i\sigma_{(0)}
\partial_j\phi_{(0)}-m^2e^{\sigma_{(0)}}\phi_{(0)} \right)~~.
\label{4.10}
\end{eqnarray}
Thus, for each energy $\omega$ one has to specify a pair of functions 
$\phi^{(0)}_\omega (x)$ and
$\psi^{(0)}_\omega (x)$ ($\phi_{(0)}(x)$ and $\psi_{(0)}(x)$ if 
$\omega=0$) on the horizon in order to
reconstruct the scalar field everywhere in the bulk. This set of functions 
forms the ``holographic data'' on the horizon.
In the AdS/CFT correspondence in order to reconstruct a scalar field in the bulk
one has to specify two functions on the boundary: a source coupled to the dual
CFT operator and the expectation value of the dual operator \cite{CMP}.
The discussion in section 2 shows that for a fixed $\omega>0$
the functions $\phi^{(0)}_\omega (x)$ and  $\phi^{(0)}_{-\omega} (x)$
(and $\psi^{(0)}_\omega (x)$ and  $\psi^{(0)}_{-\omega} (x)$) indeed form such
holographic pair.

\bigskip

\section{Reconstruction of the metric}
\setcounter{equation}0

In this section we show that provided the metric on horizon is specified 
one can uniquely reconstruct metric everywhere in bulk. 
For simplicity we consider
the static case only. The stationary and time-evolving cases are, 
however, of great importance and we plan to
approach these cases in a separate publication.

We start with the $(d+2)$-dimensional bulk Einstein equations
\begin{equation}
R_{AB}-{1\over 2}R~G_{AB}=-{d(d+1)\over 2L^2}G_{AB}~,~~A,B=(t,\rho , i)~~,
\label{5.1}
\end{equation}
where for generality we included a bulk cosmological constant 
$\Lambda\propto 1/L^2$. 
Our consideration will cover all cases:
$\Lambda>0$, $\Lambda<0$ and $\Lambda=0$. The appropriate case can be 
recovered by analytic continuation $L^2\rightarrow -L^2$ in our formulas.

Our ansatz for the metric $G_{AB}$ is as in (\ref{3.1}), i.e. 
\begin{eqnarray}
ds^2&=&e^{\sigma (x,\rho )} 
\left(-dt^2 +{{\beta_H}^2d\rho^2\over 4\rho^2}+{1\over \rho}g_{ij}(x,\rho 
)dx^idx^j \right)~~
\nonumber \\
&=&e^{\sigma (x,\rho )} \tilde{G}_{AB} dX^AdX^B~~,
\label{5.2}
\end{eqnarray}
where for convenience we separated the optical metric $\tilde{G}_{AB}$. 
The metric (\ref{5.2}) should describe a horizon at $\rho=0$, $x^i$, 
$i=1,..,d$ being the coordinates on the horizon. This is specified by
boundary conditions to be imposed on the function $\sigma (x, \rho )$. 
This, together with the analyticity
condition for the metric $g_{ij}(x, \rho )$ at $\rho=0$,  leads us to the 
following 
near-horizon expansion
\begin{eqnarray}
&&\sigma(x,\rho )=\ln \rho +\sigma_{(0)}(x)+\sigma_{(1)}(x)\rho +... 
\nonumber \\
&&g_{ij}(x,\rho )=g_{(0)\,ij}(x)+g_{(1)\,ij}(x)\rho +...
~~,
\label{5.3}
\end{eqnarray}
which we have already advocated in section 4. 

We now show that specifying the function $\sigma_{(0)}(x)$ and the metric 
$g_{(0)\,ij}(x)$ on the horizon surface $\Sigma$ uniquely determines 
the solution to the Einstein equations (\ref{5.1}) in bulk for the class 
of 
metrics (\ref{5.2}). Note also that assuming the  
expansion (\ref{5.3}) we explicitly break 
the invariance parametrized by $q$ in (\ref{3.4}).  
As we have seen in section 3 the function $\sigma_{(0)}(x)$
is pure gauge and can be always removed by using the diffeomorphism 
(\ref{3.4}) as is seen from (\ref{Weyl}).
Thus, the only real degrees of freedom living on horizon
are those of the metric function $g_{(0)\,ij}(x)$ or, equivalently, of 
the induced metric
$\gamma_{(0)\,ij}(x)$ (this is in agreement with  conclusion made  
earlier  
in \cite{SS}). However, in the rest we
will keep $\sigma_{(0)}(x)$ arbitrary in order to maintain the invariance 
under
the conformal transformation (\ref{Weyl}). Note also that 
$\sigma_{(0)}(x)$ and   $g_{(0)\,ij}(x)$
are data on the surface $\Sigma$ which is the bifurcation point in the 
horizon 
light-cone.
By means of the Killing vector $\xi_t$ these data can be extended over the 
whole 
light-cone surface.

In terms of the optical metric  $\tilde{G}_{AB}$ (\ref{5.2}) the Einstein 
equations take the form\footnote{
We are using usual conventions for the curvature tensor. Note that for 
Ricci tensor it differs in sign
from the one used in \cite{CMP}, \cite{HS}.}
\begin{eqnarray}
\tilde{R}_{AB}=-{d\over 4}(-2\tilde{\nabla}_A\tilde{\nabla}_B\sigma 
+\tilde{\nabla}_A\sigma
\tilde{\nabla}_B\sigma )+\tilde{G}_{AB} \left({d\over 
4}(\tilde{\nabla}\sigma )^2+{1\over 2}
\tilde{\nabla}^2\sigma+e^\sigma {(d+1)\over L^2}\right)~~.
\label{5.4}
\end{eqnarray}
Since we are looking for static solutions, i.e. $\partial_t 
\sigma=\partial_t g_{ij}=0$,
we find that $\tilde{R}_{tt}=0$ and 
$-2\tilde{\nabla}_t\tilde{\nabla}_t\sigma +\tilde{\nabla}_t
\sigma\tilde{\nabla}_t\sigma=0$. Hence, the $(tt)$ component of 
(\ref{5.4}) reduces to the equation
\begin{equation}
{d\over 4}(\tilde{\nabla}\sigma )^2+{1\over 2}
\tilde{\nabla}^2\sigma+e^\sigma {(d+1)\over L^2}=0~~.
\label{5.5}
\end{equation}
The remaining equations (\ref{5.4}) can be represented in the form
\begin{equation}
\tilde{R}_{\mu\nu}=-{d\over 4}\pi_{\mu\nu}~,~~\mu , \nu=(\rho , i)~~.
\label{5.6}
\end{equation}
where, in terms of $\sigma (x,\rho )$ and $g_{ij}(x, \rho )$ we have
\begin{eqnarray}
&&\pi_{\rho\rho}=-2\partial^2_\rho \sigma-{2\over \rho} \partial_\rho 
\sigma+(\partial_\rho \sigma )^2
\nonumber \\
&&\pi_{\rho i}=-2 \partial_\rho  \partial_i \sigma-{1\over \rho} 
\partial_i\sigma
+ \partial_i\sigma \partial_\rho \sigma +(g^{-1}\partial_\rho 
g)^j_i\partial_i\sigma
\nonumber \\
&&\pi_{ij}={4\over \beta_H^2}(g_{ij}-\rho\partial_\rho 
g_{ij})\partial_\rho \sigma
+ \partial_i\sigma \partial_j \sigma -2\nabla^g_i\nabla^g_j \sigma
\label{5.7}
\end{eqnarray}
and for the Ricci tensor of optical metric we have
\begin{eqnarray}
\tilde{R}_{\rho\rho}&=&-{d\over 4\rho^2}-{1\over 2}\partial_\rho \Tr 
(g^{-1}\partial_\rho g)-{1\over 4}\Tr (g^{-1}\partial_\rho g)^2 \nonumber 
\\
\tilde{R}_{\rho i}&=&-{1\over 2}\left( \partial_i \Tr (g^{-1}\partial_\rho 
g)-\nabla_j 
(g^{-1}\partial_\rho g)^j_i\right)
\nonumber \\
\tilde{R}_{ij}&=&R_{ij}[g]-\beta^{-2}_H[{d\over \rho} 
g_{ij}-(d-2)\partial_\rho g_{ij}
-g_{ij}\Tr (g^{-1}\partial_\rho g) \nonumber \\
&&+\rho \left(2\partial_\rho^2 g-2\partial_\rho g g^{-1}\partial_\rho 
g+\partial_\rho g
\Tr (g^{-1}\partial_\rho g) \right)_{ij}]~~.
\label{5.8}
\end{eqnarray}
The equations (\ref{5.6})-(\ref{5.8}) look similar to the 
$(d+1)$-dimensional Einstein equations 
analyzed in \cite{HS}, \cite{CMP} for AdS space. The essential difference 
is that $\pi_{\mu\nu}$
in (\ref{5.6}) does not represent a contribution of $(d+1)$-dimensional 
negative cosmological
constant. Only asymptotically, when $\rho \rightarrow 0$, we have that 
$-{d\over 4}\pi_{\mu\nu}=
-{d\over \beta^2_H}\tilde{G}_{\mu\nu}$, $\mu\nu=(\rho , i)$. That is why 
the solution to 
the equations (\ref{5.6})  indeed describes a 
space asymptotically approaching AdS space. However, already
in the sub-leading terms $\pi_{\mu\nu}$ deviates from the form dictated by 
the 
cosmological constant. The complete system (\ref{5.5})-(\ref{5.8}) 
describes a coupled $(d+1)$-dimensional system
of metric $g_{ij}(x,\rho )$ and a scalar field $\sigma (x,\rho )$.

Technically, we solve equations (\ref{5.5})-(\ref{5.8})
in the same way as in the case of AdS space \cite{HS}, \cite{CMP}.
One substitutes the expansion (\ref{5.3}) into the field equations which 
can then be solved order by order by matching the coefficients at fixed 
powers of $\rho$. This leads to a set of
equations sufficient to determine the coefficients $\sigma_{(k)}(x)$ and 
$g_{(k) \, ij}(x)$.
Compared to the AdS case \cite{CMP} we have one extra field $\sigma (x, 
\rho )$ but there is 
also an extra equation (\ref{5.5}). So that, the total number of equations is 
enough
to determine all unknown functions from the given holographic data.
To the leading order,  equations (\ref{5.5}), (\ref{5.6}) are satisfied 
identically. This is
a consequence
of our choice of the leading term in the expansion of the scalar function $\sigma (x,\rho )$ (\ref{5.3}).

At the first non-trivial order
eq.(\ref{5.5}) gives the relation
\begin{equation}
\sigma_{(1)}=-{1\over (d+2)}\Tr g_{(1)}-{\beta^2_H\over (d+2)}\left(
{d\over 4}({\nabla_{(0)}}\sigma_{(0)} )^2+{1\over 2}
{\nabla^2_{(0)}}\sigma_{(0)}+e^{\sigma_{(0)}} {(d+1)\over L^2}\right)
\label{5.9}
\end{equation}
between $\sigma_{(1)}$ and $g_{(1)}$.
Another relation comes from the $\rho^0$-term in the $(ij)$ component of eq. 
(\ref{5.6})
\begin{equation}
2g_{(1)\,ij}-(\Tr 
g_{(1)}+d\sigma_{(1)})g_{(0)\,ij}=\beta^2_H\left({d\over 4}(
\partial_i\sigma_{(0)}\partial_j\sigma_{(0)}-2\nabla_{(0) \, i}\nabla_{(0) \, j}
\sigma_{(0)})+R_{(0) \, ij} \right)~~.
\label{5.10}
\end{equation}
The equations (\ref{5.9}) and (\ref{5.10}) are enough to determine 
coefficients
$\sigma_{(1)}(x)$ and $g_{(1)\,ij}(x)$. After some algebra we find that
\begin{equation}
\sigma_{(1)}(x)=-{1\over 4}\beta^2_HR_{(0)}+\beta^2_H\left({d(d-3)\over 
16}(\nabla_{(0)}\sigma_{(0)})^2+
{(d-1)\over 4}\nabla_{(0)}^2\sigma_{(0)}+{(d-2)(d+1)\over 
4L^2}e^{\sigma_{(0)}}\right)
\label{5.11}
\end{equation}
\begin{eqnarray}
g_{(1)\,ij}(x)&=&{1\over 2}\beta^2_H(R_{(0) \, ij}+{1\over 
2}R_{(0)}g_{(0)\,ij})+{d\over 4}
\beta^2_H \left({1\over 2}\partial_i\sigma_{(0)}\partial_j\sigma_{(0)}
-\nabla_{(0) \, i}\nabla_{(0) \, j}\sigma_{(0)} \right) \nonumber \\
&&-{d\over 4}
\beta^2_H g_{(0)\,ij}\left({(d-1)\over 4}({\nabla_{(0)}}\sigma_{(0)} )^2+
{\nabla^2_{(0)}}\sigma_{(0)}+e^{\sigma_{(0)}} {(d+1)\over L^2}\right)~~.
\label{5.12}
\end{eqnarray}
Not all Einstein equations are independent, some give identities.
For example, the $(\rho\rho )$ part of eqs.(\ref{5.6}) does not give rise 
to any new equations on
$\sigma_{(1)}$ and $g_{(1)\,ij}$.  An identity appears from $(\rho i)$ 
part of eq.(\ref{5.6}),
\begin{equation}
\nabla_{(0)}^jg_{(1) ij}=\partial_i (\Tr g_{(1)}+d \sigma_{(1)})-
{d\over 2}(g_{(1)}+\sigma_{(1)}g_{(0)})^j_i\partial_j\sigma_{(0)}~~.
\label{5.13}
\end{equation}
As a check for our formulas one verifies that (\ref{5.13})
is indeed valid for  $\sigma_{(1)}(x)$ and $g_{(1)\,ij}(x)$ given by 
(\ref{5.11}) and (\ref{5.12}) respectively. In fact, 
the $(d+2)$-dimensional Bianchi identities imply the $(\rho i )$ equation 
to all orders provided the $(\rho\rho )$,
$(ij)$  and $(tt)$ equations are satisfied. For completeness we also give 
the 
transformation laws for $\sigma_{(1)}$ and $g_{(1)\,ij}$ under the 
remaining 
diffeomorphisms (\ref{3.4}). They are 
\begin{eqnarray}
{\cal L}_{\xi}\sigma_{(1)}(x)&=& \alpha_0(x)\sigma_{(1)}(x)-{1\over 
4}g^{ij}_{(0)}
\partial_i\alpha_0\partial_j\sigma_{(0)}
\nonumber \\
{\cal L}_{\xi}g_{(1)\,ij}(x)&=&-{1\over 2}\nabla_{(0) \, i}
\nabla_{(0) \, j} \alpha_0
 ~~.
\end{eqnarray}
In  the next order in $\rho$, the $(\rho \rho )$ part of the Einstein 
equations gives the relation
\begin{equation}
\Tr g_{(2)}={1\over 4}\Tr g^2_{(1)}+{d\over 
4}(\sigma^2_{(1)}-4\sigma_{(2)})~~.
\label{5.15}
\end{equation}
From this and the equation arising in this order from eq.(\ref{5.5}) we
can determine 
\begin{eqnarray}
\sigma_{(2)}&=&{1\over 16}\left(-3d ~ \sigma^2_{(1)}+\Tr 
g^2_{(1)}-2\sigma_{(1)}\Tr g_{(1)}
\right) 
\nonumber \\
&&-{\beta^2_H\over 16}\left(\nabla^2_{(0)}\sigma_{(1)}+{2(d+1)\over 
L^2}e^{\sigma_{(0)}}\sigma_{(1)}\right)
+\dots~~,
\label{5.16}
\end{eqnarray}
where the ellipses represent terms involving derivatives of $\sigma_{(0)}$ 
which can be set to zero by a suitable diffeomorphism of the form 
(\ref{3.4}). Then, from  $(ij)$ part of eqs.(\ref{5.6}) we determine 
\begin{eqnarray}
g_{(2) \, ij}&=&{1\over 16}\left(4g^2_{(1)}+g_{(0)}(d \, \sigma^2_{(1)}-\Tr 
g^2_{(1)})\right)_{ij}
-{\beta^2_H\over 16}\left(d \, \nabla_{(0) \, i}\nabla_{(0) \, j}\sigma_{(1)} 
+\nabla_{(0)\, i}\nabla_{(0) \, j}
\Tr g_{(1)}\right) \nonumber \\
&&-{\beta^2_H\over 
16}\left(\nabla^2_{(0)}g_{(1)\,ij}-\nabla_{(0)}^k\nabla_{(0) \, i} ~
g_{(1)\,kj}
-\nabla_{(0)}^k\nabla_{(0) \, j} ~ g_{(1)\,ki}\right)+\dots~~.
\label{5.17}
\end{eqnarray}
In higher orders the strategy remains the same but the expressions become 
more complicated. However,
the general structure of the equations on coefficients $\sigma_{(k)}$ and
$g_{(k) \, ij}$ can be easily analyzed. From eq.(\ref{5.5}) we find that
$$
(d+2k)\sigma_{(k)}+\Tr g_{(k)}=A_{(k)}[\sigma_{(0)},g_{(0)},..., 
\sigma_{(k-1)},g_{(k-1)}]~~.
$$
The ($\rho\rho )$ equation leads to
$$
\Tr g_{(k)}+d\sigma_{(k)}=B_{(k)}[\sigma_{(0)},g_{(0)},..., 
\sigma_{(k-1)},g_{(k-1)}]~~.
$$
$A_{(k)}$ and $B_{(k)}$ are some functions of the coefficients 
$\sigma_{(i)}$, $g_{(i)}$, $i<k$
and their derivatives. These equations determine $\sigma_{(k)}$ and $\Tr 
g_{(k)}$ in terms of
$\sigma_{(i)}$, $g_{(i)}$, $i<k$.
The $(ij)$ equation  takes the form
$$
(2k g_{(k)}-g_{(0)}\Tr g_{(k)})_{ij}=C_{(k)ij}[\sigma_{(0)},g_{(0)},..., 
\sigma_{(k-1)},g_{(k-1)}]~~
$$
and determines the coefficient $g_{(k)ij}$. We see that no ambiguity 
arises 
in this iteration process and all coefficients in the expansion 
(\ref{5.3}) are uniquely determined
once $\sigma_{(0)}(x)$ and $g_{(0)\,ij}(x)$ are specified on horizon.

Reconstructing the metric in the AdS/CFT correspondence \cite{CMP}
one has to specify
the metric on the boundary and the expectation value of the stress tensor
of dual CFT. Therefore, two terms in the expansion in $\rho$ of the bulk metric 
remain undetermined (for more details see \cite{CMP}): 
the metric on the boundary $g_{(0) \, ij}(x)$
and the term $g_{(2d) \, ij}(x)$ related to the value of the extrinsic curvature on 
the boundary. One might have expected the same 
to happen in our case. However, surprisingly only
one tensor $g_{(0) \, ij}(x)$ needs to be specified, all the other terms in the expansion
are then uniquely determined. This is due to the peculiar 
properties of the horizon (see also \cite{Dzhu} for the spherically 
symmetric case). In particular, the $(ij)$ components of the 
extrinsic curvature vanish on the horizon.

\bigskip

\section{Non-spherical horizons, uniqueness and entropy}
\setcounter{equation}0

The results obtained in the previous section may be interesting for
conventional General Relativity. What we have found is solution to 
Einstein equations describing a
space-time with a horizon with arbitrary (not necessarily spherically 
symmetric)
metric. It is worth reminding that in General Relativity the possible 
shape of the horizon is constrained 
by the so-called Uniqueness Theorem proven by W. Israel \cite{WI} (for a 
review see
\cite{REV}).
According to this Theorem if the space-time metric is

1) asymptotically flat;

2) has an event horizon;

3) has no singularity on or outside the event horizon;

4) satisfies the vacuum Einstein equations

\noindent then it is spherically symmetric and in empty space coincides 
with the 
Schwarzschild metric.

In section 5 we gave up the condition 1) since the solution 
we have found is known explicitly only
in vicinity of the horizon and apparently can not be 
asymptotically flat at infinity. Also, for the Ansatz (\ref{3.1}) we can 
not 
rule out singularities outside the horizon.  
The only  exception from the Theorem known before is the so-called 
topological black holes
\cite{TB} (for a review see \cite{REV2}).
They are solutions to Einstein equations with cosmological 
constant.
The horizon in this case is a hypersurface of constant curvature 
which, 
after some identifications, can be made compact and topologically 
non-trivial.
However, in this case the horizon is still a maximally symmetric space. In 
the solution found in
section 5 the metric on horizon can be arbitrary and it does not require 
any bulk
cosmological constant for its existence.
So that it is a new class of metrics describing space-times with horizon.
It may find applications, for instance, in the analysis of fluctuations of 
the black hole horizon.

Finally, one may wonder whether it is meaningful to discuss 
thermodynamical properties of the configurations discussed in this paper. 
For instance, what is the horizon entropy in this case?
A standard method to determine the entropy is based on the first law
and is not applicable  in our case since the solution is known explicitly 
only 
locally near horizon. Therefore, its global characteristics like mass are 
not
determined.

The method which does work in this case is the method of conical 
singularity \cite{FS}. 
It consists in 
proceeding along the following steps

1. go to Euclidean signature, $\tau=\imath t$

2. close Euclidean time $\tau$ with arbitrary period $2\pi \beta$, 
$\beta\neq \beta_H$.
Then there appears a conical singularity at horizon surface $\Sigma$ with 
angle deficit
$\delta=2\pi (1-\alpha )$, $\alpha=\beta /\beta_H$. The Ricci scalar on 
such conical 
manifold has a contribution from the singularity \cite{FS}
\begin{equation}
R=4\pi (1-\alpha ) \delta_\Sigma +R_{\alpha=1}~~,
\label{6.1}
\end{equation}
where $R_{\alpha=1}$ is the regular part of the curvature.

3. compute the Einstein-Hilbert action $W_{EH}[M_\alpha ]=-{(1-\alpha 
)\over 4G}A_\Sigma -
{\alpha \over 16\pi G}\int_{M_{\alpha=1}} R_{\alpha=1}$ on the conical 
space and apply the formula
\begin{equation}
S=(\alpha\partial_\alpha -1 )W_{EH}[M_{\alpha}]|_{\alpha=1}~~
\label{6.2}
\end{equation}
to compute the entropy.
Since the entropy comes entirely as a contribution of the conical 
singularity, only 
the metric near the horizon is essential for the computing the entropy in this 
method.

The analysis of paper \cite{FS} is quite general and requires only the 
knowing the metric
in the form (\ref{5.2}) near horizon as an  expansion (\ref{5.3})\footnote{
In fact, the results of \cite{FS} are even more general since they are 
still valid when
the coefficients in the expansion (\ref{5.3}) are functions of Euclidean 
time $\tau$.}. 
With this definition, the entropy is found to be universally proportional 
to the horizon area
$A_\Sigma=\int_\Sigma d^dx \sqrt{\gamma_{(0)}}$
\begin{equation}
S_{BH}={A_\Sigma\over 4G}~~,
\label{6.3}
\end{equation}
independently on the shape or topology of the horizon surface $\Sigma$.

An alternative method for computing the entropy is Wald's method of 
Noether charges \cite{Noether}.
It would be interesting to see if this method applies to the present 
situation.

\bigskip

\section{Conclusions}
\setcounter{equation}0

The purpose of this paper was to initiate a systematic study of 
holographic encoding on a static, non-degenerate, but 
otherwise arbitrary horizon. Our basic observation is the presence of a 
universal near horizon structure conformally related to 
${{\rm I\!R}}\times$ Euclidean anti-de Sitter space. 
This is best described in terms of the optical 
metric. Moreover by analyzing the diffeomorphisms compatible with this 
near horizon structure we revealed the presence of an asymptotic conformal 
symmetry. We also computed the $2$-point functions for a bulk scalar 
perturbation and found agreement with the expected form for some conformal 
field theory located on the horizon. Although, in this paper we have 
concentrated on static space-times it is conceivable that our results can 
be extended to more general space-times. 

In spite of these similarities with the AdS/CFT 
correspondence there are also important differences. 
Notably we find that the conformal weights are complex. 
In section $3$ we made some speculations 
about possible interpretations of this phenomenon, but other 
interpretations are possible and it would be interesting to learn more 
about the CFT side and to what extent the analogy with the AdS/CFT 
correspondence is complete. In particular, it would be interesting see if the 
horizon entropy and Hawking radiation has a CFT representation in this case. 

Finally we showed that the metric as well as scalar fields can be 
unambiguously reconstructed from the horizon data by an iterative process 
in much the same way as in the anti-de Sitter case.

\bigskip

\section*{Acknowledgements}

We would like to thank A. Barvinsky for helpful 
discussions. This work has been supported by the 
DFG-Stringtheorie Schwerpunktsprogramm SPP 1096.


\begin{thebibliography} \\
\bibitem{tHooft}
G. 't Hooft, ``Dimensional reduction in Quantum Gravity'', in 
Salamfestschrift: A Collection of Talks, World Scientific Series in 20th 
Century Physics, Vol. 4, ed. A. Ali, J. Ellis and S. Randjbar-Daemi (World 
Scientific, 1993), 
gr-qc/9310026.

\bibitem{Susskind} L. Susskind, ``The World as a Hologram'', 
J. Math. Phys. {\bf 36} (1995) 6377, hep-th/9409089.


\bibitem{Malda} J. Maldacena, ``The Large N Limit of Superconformal
Field Theories and Supergravity'', 
Adv. Theor. Math. Phys. {\bf 2} (1998) 231,
hep-th/9711200.

\bibitem{Gubs} S. Gubser, I. Klebanov and A. Polyakov,
``Gauge Theory Correlators from Non-Critical String Theory'',
Phys. Lett. {\bf B428} (1998) 105,
hep-th/9802109.


\bibitem{Wit} E. Witten, ``Anti De Sitter Space And Holography'',
Adv. Theor. Math. Phys. {\bf 2} (1998) 253, hep-th/9802150.


\bibitem{CMP} S. de Haro, K. Skenderis and S. N. Solodukhin, ``Holographic 
Reconstruction of Space-time and Renormalization in the AdS/CFT 
Correspondence'',
Commun. Math. Phys. {\bf 217} (2001) 595, hep-th/0002230


\bibitem{BH} J. D. Bekenstein, Phys.Rev. {\bf D7} (1973) 2333; 
S. W. Hawking, Comm. Math. Phys. {\bf 43} (1975) 199.


\bibitem{Carlip0} S. Carlip, ``The Statistical Mechanics of the 
(2+1)-Dimensional Black Hole'', 
Phys. Rev. {\bf D51} (1995) 632, gr-qc/9409052.  
     
\bibitem{Carlip} S. Carlip, ``Black hole entropy from conformal field 
theory in any dimension'',
Phys. Rev. Lett.  {\bf 82} (1999) 2828, hep-th/9812013.

\bibitem{SS} S. N. Solodukhin, ``Conformal description of horizon's 
states'',
Phys. Lett. {\bf B454} (1999), 213, hep-th/9812056.


\bibitem{FS} D. V. Fursaev and S. N. Solodukhin, ``On the description of 
the Riemannian geometry in the presence
of conical defects'', Phys. Rev. {\bf D52} (1995), 2133.

\bibitem{Dzhu} V. Dzhunushaliev, "Matching Condition on the Event Horizon 
and Hologhraphy", gr-qc/9907086. 

\bibitem{SS1} S. N. Solodukhin, ``Correlation functions of boundary field 
theory from
bulk Green's functions and phases in the boundary theory'', Nucl. Phys. 
{\bf B539} (1999) 403, hep-th/9806004.

\bibitem{GtH} G. 't Hooft, ``Graviton Dominance in Ultra-High-Energy 
Scattering'',
Phys. Lett. {\bf B198} (1987) 61; Nucl. Phys. {\bf B335} (1990) 138.

\bibitem{VV} E. Verlinde and H. Verlinde, ``Scattering at Planckian 
Energies'',
Nucl. Phys. {\bf B371} (1992) 246.


\bibitem{Wit1} E. Witten, ``Quantum Gravity In De Sitter Space'', 
hep-th/0106109.

\bibitem{Str}  A. Strominger, ``The dS/CFT Correspondence'', 
hep-th/0106113.

\bibitem{FG} C. Fefferman and C. R. Graham: ``Conformal Invariants''. In: 
{\it Elie Cartan et les Mathematiques d'aujord'hui}, (Asterisque, 1985), 
95.

\bibitem{GL} C. R. Graham and J. M. Lee,  ``Einstein metrics with 
prescribed conformal infinity on the
ball'', Adv. Math. {\bf 87} (1991) 186.

\bibitem{HS} M. Henningson and K. Skenderis, ``The holographic Weyl 
anomaly'', JHEP {\bf 9807} (1998), 023,
hep-th/9806087; ``Holography and the Weyl anomaly'', hep-th/9812023.

\bibitem{Brown} J. D. Brown and M. Henneaux, 
``Central charges in the canonical realization of asymptotic symmetries: 
an example from three dimensional gravity'', Commun. Math. Phys. {\bf 104}
 (1986) 207. 

\bibitem{ISTY} C. Imbimbo, A. Schwimmer, S. Theisen and S. Yankielowicz, 
``Diffeomorphisms and Holographic 
Anomalies'', Class. Quant. Grav. {\bf 17} (1999) 1129, hep-th/9910267.

\bibitem{WI} W. Israel, ``Event horizons in static vacuum space-times'',
Phys. Rev. {\bf 164}  (1967) 1776.

\bibitem{REV} J. D. Bekenstein, ``Black hole hair: twenty-five years 
after'', gr-qc/9605059.

\bibitem{TB} S. Aminneborg, I. Bengtsson, S. Holst and P. Peldan, Class. 
Quant. Grav.
{\bf 13} (1996) 2707; D. Brill, Helv. Phys. Acta {\bf 69} (1996) 249;
M. Banados, ``Black holes of constant curvature'', gr-qc/9703040; 
D. Birmingham, "Topological Black Holes in Anti-de Sitter Space", 
Class. Quant. Grav. {\bf 16} (1999) 1197, hep-th/9808032. 

\bibitem{REV2} R. B. Mann, ``Topological black holes - outside looking 
in'', gr-qc/9709039.

\bibitem{Noether}  R. M. Wald, Phys. Rev. {\bf D48} (1993) R3427; V. Iyer 
and R. M. Wald, 
Phys. Rev. {\bf D50} (1994) 846. T. A. Jacobson, G. Kang and R. C. Myers,
Phys. Rev. {\bf D49} (1994) 6587.

\end{thebibliography}
\end{document}